\documentclass[a4paper,1pt]{article}

\usepackage{fancyhdr}
\usepackage{multicol}
\usepackage{etoolbox}
\usepackage{cancel}
\usepackage[table]{xcolor}
 
\usepackage[font=small,labelfont=bf]{caption}
\DeclareCaptionFont{scriptsize}{\scriptsize}
\DeclareCaptionFont{tiny}{\tiny}
\usepackage{float}
\usepackage{amsmath, amssymb, setspace,cite,verbatim}
\usepackage{graphicx}
\usepackage{hyperref}
\usepackage{color}
\usepackage{wrapfig}

\usepackage{multicol}
\usepackage{etoolbox}
\usepackage{relsize}
\usepackage{multirow}

\usepackage{latexsym}
\usepackage{a4wide}
\usepackage{graphicx, subfigure}
\usepackage{cite}
\usepackage{tabularx}
\usepackage{array}
\usepackage{graphics}     
\usepackage{amsmath}
\usepackage{amssymb}
\usepackage{longtable} 
\usepackage{verbatim} 
\usepackage[table]{xcolor}
\usepackage{booktabs}
\usepackage{hyperref} 
\usepackage[utf8]{inputenc}
\usepackage{soul}
\usepackage[table]{xcolor}
\usepackage[normalem]{ulem}

\usepackage{pifont}

\definecolor{light-gray}{gray}{0.90}

\begin{document}


\renewcommand*{\thefootnote}{\fnsymbol{footnote}}

\begin{center}

{\large\bf{Model independent analysis of the angular observables}}\\ 
\vspace{0.3cm}
{\large\bf{in $B^{0} \to K^{*0} \mu^+ \mu^-$ and $B^{+} \to K^{*+} \mu^+ \mu^-$~\footnote{Addendum to Phys. Rev. D102 (2020) 055001, arXiv:2006.04213.}}}\\

\setlength{\textwidth}{11cm}
                    
\vspace{0.7cm}
{
T.~Hurth$^{a,}$\footnote{Email: tobias.hurth@cern.ch},
F.~Mahmoudi$^{b,c,}$\footnote{Email: nazila@cern.ch},
S.~Neshatpour$^{b,}$\footnote{Email: siavash.neshatpour@univ-lyon1.fr }
}

\vspace{0.3cm} 
{\small {\em $^a$PRISMA+ Cluster of Excellence and Institute for Physics (THEP)\\
Johannes Gutenberg University, D-55099 Mainz, Germany}\\
{\em $^b$Universit\'e de Lyon, Universit\'e Claude Bernard Lyon 1, CNRS/IN2P3, \\
Institut de Physique des 2 Infinis de Lyon, UMR 5822, F-69622, Villeurbanne, France}\\
{\em $^c$Theoretical Physics Department, CERN, CH-1211 Geneva 23, Switzerland} \\
}
\end{center}

\renewcommand{\thefootnote}{\arabic{footnote}}
\setcounter{footnote}{0}
\vspace{0.8cm}
{\bf Abstract:} We analyse the results recently presented on the 
$B^{+} \to K^{*+} \mu^+ \mu^-$ angular observables by the LHCb Collaboration which show indications for New Physics beyond the Standard Model.  
Within a model-independent analysis, we compare the fit results with the corresponding results for the angular observables in $B^{0} \to K^{*0} \mu^+ \mu^-$. \\[0.5cm]

Following the latest update of the angular analysis of 
the decay $B^{0} \to K^{*0} \mu^+ \mu^-$~\cite{Aaij:2020nrf} in March 2020, the LHCb collaboration has very recently presented a new angular analysis in the $B^{+} \to K^{*+} \mu^+ \mu^-$ decay~\cite{Aaij:2020ruw}.  
In particular, for the first time the observable $P_5^{'}$ is measured outside the $B^{0}$ meson system. Both analyses show some tensions with the Standard Model (SM) predictions. 
It is very suggestive to compare the tensions in these two modes using model-independent methods in order to check their consistency. 
Theoretically, the two modes are connected by isospin symmetry and no large differences are expected. Here we update and compare our previous results in Ref.~\cite{Hurth:2020rzx} with new fits for the $B^{+}$ mode. See also Ref.~\cite{Ciuchini:2020gvn} for a recent global fit including $B^+\to K^{*+} \mu^+ \mu^-$ observables.

In Table~\ref{tab:BorBplus_1D_A}, we compare one-operator 
fits using the two sets of angular observables. 
As in our previous analyses we assume $10\%$ uncertainty for the power corrections 
(see Refs.~\cite{Arbey:2018ics,Hurth:2016fbr} for more details). 
Clearly, the SM pull in the $B^{0}$ mode is significantly  
larger due to much smaller experimental errors, which are on average about a factor five smaller than in the $B^{+}$ mode.

The second major difference is that the best fit values differ in all cases, preferring larger New Physics (NP) contributions in the $B^{+}$ mode. 
For $\delta C_7$ and $\delta C_{10}^\mu$ one must keep in mind 
that the constraints of $\bar B \to X_s \gamma$ 
and $B_{s,d} \to \mu^+\mu^-$ are not taken into account here. 
In the case of $\delta C_{10}^\mu$, negative values are preferred, while we know that the ratios $R_{K^{(*)}}$, indicating flavour non-universality, ask for positive NP contributions. For the $\delta C_9^\mu$ case we find in the recent $B^{+}$ mode a best fit value which is larger by a factor two as compared to the one in the $B^{0}$ mode.

However, regarding NP significance the hierarchy among the various one-parameter fits is the same in both modes. 
It is again the coefficient $C_9^\mu$ which is favoured by both sets of data.  
This consistency is also manifest when we fit for the chiral Wilson coefficients, 
given in Table~\ref{tab:BorBplus_1D_B}.
Thus, at the level of the one-operator fits the two sets guide us to the same NP patterns.

\begin{table}[th!]
\begin{center}
\hspace*{-1.cm}
\scalebox{0.90}{
\qquad\quad
\begin{tabular}{|l|r|r|c|}
\hline 
\multicolumn{4}{|c|}{$B^0 \to K^{*0} \mu^+ \mu^-$ ang. obs.\;  ($\chi^2_{\rm SM}=	81.86	$)} \\ \hline								
& b.f. value & $\chi^2_{\rm min}$ & ${\rm Pull}_{\rm SM}$  \\										
\hline \hline										
$\delta C_7$	& $ 	-0.11	\pm	0.02	 $ & $ 	61.4	 $ & $	4.5	\sigma	 $  \\
$\delta C_{9}^{\mu} $    	& $ 	-1.03	\pm	0.15	 $ & $ 	52.8	 $ & $	5.4	\sigma	 $  \\
$\delta C_{10}^{\mu} $    	& $ 	-1.20	\pm	0.50	 $ & $ 	75.0	 $ & $	2.6	\sigma	 $  \\
\hline							          			
\end{tabular}\qquad\qquad\qquad
\begin{tabular}{|l|r|r|c|}
\hline 
\multicolumn{4}{|c|}{$B^+ \to K^{*+} \mu^+ \mu^-$ ang. obs.\;  ($\chi^2_{\rm SM}=	58.52	$)} \\ \hline								
& b.f. value & $\chi^2_{\rm min}$ & ${\rm Pull}_{\rm SM}$  \\										
\hline \hline										
$\delta C_7$	& $ 	-0.27	\pm	0.09	 $ & $ 	49.4	 $ & $	3.0	\sigma	 $  \\
$\delta C_{9}^{\mu} $    	& $ 	-2.06	\pm	0.34	 $ & $ 	43.3	 $ & $	3.9	\sigma	 $  \\
$\delta C_{10}^{\mu} $    	& $ 	-4.70	\pm	2.10	 $ & $ 	51.4	 $ & $	2.7	\sigma	 $  \\
\hline							          			
\end{tabular}
}
\caption{One-operator fits to only $B^0 \to K^{*0} \mu^+ \mu^-$ observables on the left and
to only $B^+ \to K^{*+} \mu^+ \mu^-$ observables on the right, considering 
10\% power corrections for both cases.
\label{tab:BorBplus_1D_A}} 
\vspace{1cm}

\scalebox{0.83}{
\begin{tabular}{|l|r|r|c|}
\hline 
\multicolumn{4}{|c|}{$B^0 \to K^{*0} \mu^+ \mu^-$ ang. obs.\;  ($\chi^2_{\rm SM}=	81.86	$)} \\ \hline								
& b.f. value & $\chi^2_{\rm min}$ & ${\rm Pull}_{\rm SM}$  \\										
\hline \hline										
$\delta C_{\rm LL}^\mu$ \,  ($\delta C_{9}^{\mu}=-\delta C_{10}^{\mu}$)           	& $ 	-0.93	\pm	0.17	 $ & $ 	63.8	 $ & $	4.3	\sigma	 $  \\
$\delta C_{\rm LR}^\mu$ \,  ($\delta C_{9}^{\mu}=+\delta C_{10}^{\mu}$)	& $ 	-0.76	\pm	0.13	 $ & $ 	56.3	 $ & $	5.1	\sigma	 $  \\
$\delta C_{\rm RL}^\mu$ \,  ($\delta C_{9}^{\mu \prime}=-\delta C_{10}^{\mu \prime}$)	& $ 	-0.35	\pm	0.16	 $ & $ 	77.1	 $ & $	2.2	\sigma	 $  \\
$\delta C_{\rm RR}^\mu$ \,  ($\delta C_{9}^{\mu \prime}=+\delta C_{10}^{\mu \prime}$)	& $ 	0.51	\pm	0.26	 $ & $ 	77.8	 $ & $	2.0	\sigma	 $  \\
\hline
\end{tabular}
}\quad
\scalebox{0.835}{
\begin{tabular}{|l|r|r|c|}
\hline 
\multicolumn{4}{|c|}{$B^+ \to K^{*+} \mu^+ \mu^-$ ang. obs.\;  ($\chi^2_{\rm SM}=	58.52	$)} \\ \hline								
& b.f. value & $\chi^2_{\rm min}$ & ${\rm Pull}_{\rm SM}$  \\										
\hline \hline										
$\delta C_{\rm LL}^\mu$ \,  ($\delta C_{9}^{\mu}=-\delta C_{10}^{\mu}$)           	& $ 	-1.80	\pm	0.40	 $ & $ 	51.0	 $ & $	2.8	\sigma	 $  \\
$\delta C_{\rm LR}^\mu$ \,  ($\delta C_{9}^{\mu}=+\delta C_{10}^{\mu}$)	& $ 	-1.70	\pm	0.40	 $ & $ 	43.6	 $ & $	3.9	\sigma	 $  \\
$\delta C_{\rm RL}^\mu$ \,  ($\delta C_{9}^{\mu \prime}=-\delta C_{10}^{\mu \prime}$)	& $ 	-1.40	\pm	0.50	 $ & $ 	48.1	 $ & $	3.2	\sigma	 $  \\
$\delta C_{\rm RR}^\mu$ \,  ($\delta C_{9}^{\mu \prime}=+\delta C_{10}^{\mu \prime}$)	& $ 	1.00	\pm	0.80	 $ & $ 	56.9	 $ & $	1.3	\sigma	 $  \\
\hline
\end{tabular}
}
\caption{One-operator fits in the chiral basis using only $B^0 \to K^{*0} \mu^+ \mu^-$ observables on the left and
only $B^+ \to K^{*+} \mu^+ \mu^-$ observables on the right, considering 
10\% power corrections for both cases.
\label{tab:BorBplus_1D_B}}
\end{center} 
\end{table}
%

\begin{table}[th!]
\begin{center}
\hspace*{-1.cm}
\scalebox{0.9}{
\qquad\quad
\begin{tabular}{|c||c|c||c|c|}
\hline 
\multicolumn{5}{|c|}{ $B^0 \to K^{*0} \mu^+ \mu^-$ ang. obs.} \\ \hline
 & \multicolumn{2}{c||}{excluding $[6,8]$ GeV$^2$ bin} & \multicolumn{2}{c|}{including $[6,8]$ GeV$^2$ bin}\\ \hline
$\delta C_{9}^{\mu} $  & b.f. value & ${\rm Pull}_{\rm SM}$  & b.f. value & ${\rm Pull}_{\rm SM}$  \\						\hline \hline										
SuperIso 	& $ -0.86 \pm 0.19 $ 	& $ 3.8\sigma $ & $ -1.05 \pm 0.15 $ 	& $ 5.4\sigma $ \\
LHCb 		& $ -   $ 		& $ 2.7\sigma $ & $ -0.99 \pm 0.25 $ 	& $ 3.3\sigma $ \\
\hline							          			
\end{tabular}\qquad
\begin{tabular}{|c||c|c|}
\hline 
\multicolumn{3}{|c|}{$B^+ \to K^{*+} \mu^+ \mu^-$ ang. obs.}\\ \hline
& \multicolumn{2}{c|}{excluding $[6,8]$ GeV$^2$ bin} \\ \hline
$\delta C_{9}^{\mu} $  	& b.f. value 		& ${\rm Pull}_{\rm SM}$    \\ \hline \hline										
SuperIso 		& $ -1.80 \pm 0.40 $ 	& $ 2.9\sigma $ \\
LHCb 			& $ -1.90 $ 		& $ 3.1\sigma	 $ \\
\hline							          			
\end{tabular}
}
\caption{One-dimensional fit to $C_9^\mu$ by 
SuperIso and LHCb, considering the angular observables 
of either the $B^0 \to K^{*0} \mu^+ \mu^-$ decay or the $B^+ \to K^{*+} \mu^+ \mu^-$ decay (the best fit value  
for the $B^0$ mode when excluding the $[6,8]$ GeV$^2$ bin, and the error on the best fit value for the $B^+$ mode was not given by LHCb). 
\label{tab:SuperIso_vs_LHCb}}
\end{center} 
\end{table}

In Table~\ref{tab:SuperIso_vs_LHCb}, we directly compare the NP significance that we obtain for $C_9^\mu$ with the NP significance LHCb quotes in their analysis of the $B^{0}$ mode.  The drastic difference, $5.4 \sigma$ versus $3.3 \sigma$,  
reflects the fact that LHCb uses the Flavio package~\cite{Straub:2018kue} 
while in the present analysis SuperIso~\cite{Mahmoudi:2007vz} is used.
SuperIso and Flavio use the same set of form factors and 
similar input parameters, but there are differences in the parameterisation of the power corrections. In SuperIso 
we consider $10\%$  on top of the leading order non-local effects (known 
from QCDf calculations), motivated by the fact that it is only higher powers of these contributions that are to be accounted for, while in Flavio the additional power correction uncertainty is considered on both factorisable and non-factorisable pieces (multiplied by $C_7^{\rm eff}$ or $C_9^{\rm eff}$), which inflates the errors artificially. 
Hence even with similarly assumed percentages the two parameterisations  
lead to the different NP significances.
The different assumptions about the power corrections have no impact in the $B^{+} $ mode because the large experimental uncertainties dominate
in this mode. Furthermore, in view of the recent theoretical analyses~\cite{Bobeth:2017vxj,Chrzaszcz:2018yza,Gubernari:2020eft} the assumptions on power corrections in SuperIso seem to be more realistic.  

In the global fit to all relevant $b \to s$ observables including the recent $B^+$ mode,
the hierarchy of the favoured one-dimensional NP scenarios has remained the same with
$\delta C_9^{\mu}$ as the most favoured scenario. In Table~\ref{tab:ALLwBplus_1D}  we present the full one-operator fit results using the complete set of $b \to s$ observables described in Ref.~\cite{Hurth:2020rzx} and the new $B^+ \to K^{*+} \mu^+ \mu^-$ angular observables. 

It is more reasonable to assume that a UV complete NP model affects several Wilson coefficients at the same time. 
Therefore, we consider the NP significance of a multi-dimensional 
fit using the full set of 20 Wilson coefficients. The results are given in Table~\ref{tab:ALLwBplus_20D_C78910C12primes}. 
In Ref.~\cite{Hurth:2020rzx} we found 
that the update of the $B^{0}$ dataset changed the global fit significantly (by $0.8\sigma$) compared to the previous global fits (see Table 8 in Ref.~\cite{Arbey:2018ics}). 
Surprisingly, this is also true when including the $B^{+}$ data. 
Adding the new data on the $B^{+}$ mode to the global fit we find an additional increase of the NP significance by $0.5 \sigma$ in spite of the comparably larger experimental error. However, as shown above, the new data on the $B^{+}$ mode leads also to a best fit value for the 
favoured NP Wilson coefficient $\delta C_9^\mu$ which is different from the best fit value of the 
same coefficient in the global fit by a factor two.\\

\begin{table}[th!]
\begin{center}
\setlength\extrarowheight{3pt}
\hspace*{-1.cm}
\scalebox{0.92}{\qquad
\begin{tabular}{|l|r|r|c|}
\hline 
\multicolumn{4}{|c|}{All observables  ($\chi^2_{\rm SM}=	215.1	$)} \\ \hline								
& b.f. value & $\chi^2_{\rm min}$ & ${\rm Pull}_{\rm SM}$  \\										
\hline \hline										
$\delta C_7$	& $ 	-0.03	\pm	0.01	 $ & $ 	207.3	 $ & $	2.8	\sigma	 $  \\
$\delta C_{9} $    	& $ 	-1.01	\pm	0.13	 $ & $ 	177.4	 $ & $	6.1	\sigma	 $  \\
$\delta C_{9}^{e} $    	& $ 	0.84	\pm	0.26	 $ & $ 	203.4	 $ & $	3.4	\sigma	 $  \\
$\delta C_{9}^{\mu} $    	& $ 	-0.99	\pm	0.12	 $ & $ 	165.9	 $ & $	7.0	\sigma	 $  \\
$\delta C_{10} $    	& $ 	0.16	\pm	0.21	 $ & $ 	214.5	 $ & $	0.8	\sigma	 $  \\
$\delta C_{10}^{e} $    	& $ 	-0.79	\pm	0.23	 $ & $ 	202.1	 $ & $	3.6	\sigma	 $  \\
$\delta C_{10}^{\mu} $    	& $ 	0.50	\pm	0.17	 $ & $ 	205.1	 $ & $	3.2	\sigma	 $  \\
\hline							          			
\end{tabular}
} \qquad 
\scalebox{0.82}{
\begin{tabular}{|l|r|r|c|}
\hline 
\multicolumn{4}{|c|}{All observables  ($\chi^2_{\rm SM}=	215.1	$)} \\ \hline								
& b.f. value & $\chi^2_{\rm min}$ & ${\rm Pull}_{\rm SM}$  \\										
\hline \hline										
$\delta C_{\rm LL}^e$ \,  ($\delta C_{9}^{e}=-\delta C_{10}^{e}$)	& $ 	0.43	\pm	0.14	 $ & $ 	202.7	 $ & $	3.5	\sigma	 $  \\
$\delta C_{\rm LL}^\mu$ \,  ($\delta C_{9}^{\mu}=-\delta C_{10}^{\mu}$)           	& $ 	-0.55	\pm	0.10	 $ & $ 	180.7	 $ & $	5.9	\sigma	 $  \\
\hline										
$\delta C_{\rm LR}^e$ \,  ($\delta C_{9}^{e}=+\delta C_{10}^{e}$)	& $ 	-1.64	\pm	0.29	 $ & $ 	201.1	 $ & $	3.7	\sigma	 $  \\
$\delta C_{\rm LR}^\mu$ \,  ($\delta C_{9}^{\mu}=+\delta C_{10}^{\mu}$)	& $ 	-0.43	\pm	0.11	 $ & $ 	203.5	 $ & $	3.4	\sigma	 $  \\
\hline										
$\delta C_{\rm RL}^e$ \,  ($\delta C_{9}^{e \prime}=-\delta C_{10}^{e \prime}$)	& $ 	0.07	\pm	0.11	 $ & $ 	214.7	 $ & $	0.6	\sigma	 $  \\
$\delta C_{\rm RL}^\mu$ \,  ($\delta C_{9}^{\mu \prime}=-\delta C_{10}^{\mu \prime}$)	& $ 	-0.08	\pm	0.07	 $ & $ 	213.8	 $ & $	1.1	\sigma	 $  \\
\hline										
$\delta C_{\rm RR}^e$ \,  ($\delta C_{9}^{e \prime}=+\delta C_{10}^{e \prime}$)	& $ 	1.81	\pm	0.30	 $ & $ 	202.0	 $ & $	3.6	\sigma	 $  \\
$\delta C_{\rm RR}^\mu$ \,  ($\delta C_{9}^{\mu \prime}=+\delta C_{10}^{\mu \prime}$)	& $ 	0.14	\pm	0.14	 $ & $ 	214.2	 $ & $	0.9	\sigma	 $  \\
\hline
\end{tabular}
}
\caption{One-operator fits to all observables, assuming 10\% error for the power corrections. 
\label{tab:ALLwBplus_1D}} 
\end{center} 
\end{table}
%
\begin{table}[th!]
\begin{center}
\setlength\extrarowheight{3pt}
\scalebox{0.78}{
\begin{tabular}{|c|c|c|c|}
\hline																
\multicolumn{4}{|c|}{All observables (excl. $B^+ \to K^{*+} \mu^+ \mu^-$)  with $\chi^2_{\rm SM}=	 	157.5			$} \\											
\multicolumn{4}{|c|}{($\chi^2_{\rm min}=	 	100.4	;\; {\rm Pull}_{\rm SM}=	4.3	\sigma$)} \\											
\hline \hline																
\multicolumn{2}{|c|}{$\delta C_7$} &  \multicolumn{2}{c|}{$\delta C_8$}\\																
\multicolumn{2}{|c|}{$	0.05	\pm	0.03	$} & \multicolumn{2}{c|}{$	-0.70	\pm	0.40	$}\\								
\hline																
\multicolumn{2}{|c|}{$\delta C_7^\prime$} &  \multicolumn{2}{c|}{$\delta C_8^\prime$}\\																
\multicolumn{2}{|c|}{$	-0.01	\pm	0.02	$} & \multicolumn{2}{c|}{$	-0.10	\pm	0.80	$}\\								
\hline																
$\delta C_{9}^{\mu}$ & $\delta C_{9}^{e}$ & $\delta C_{10}^{\mu}$ & $\delta C_{10}^{e}$ \\																
$	-1.12	\pm	0.19	$ & $	-6.70	\pm	0.90	$ & $	0.12	\pm	0.23	$ & $	4.00	\pm	1.70	$ \\
\hline\hline																
$\delta C_{9}^{\prime \mu}$ & $\delta C_{9}^{\prime e}$ & $\delta C_{10}^{\prime \mu}$ & $\delta C_{10}^{\prime e}$ \\																
$	0.13	\pm	0.33	$ & $	1.80	\pm	1.10	$ & $	-0.12	\pm	0.21	$ & $	0.00	\pm	2.00	$ \\
\hline\hline																
$C_{Q_{1}}^{\mu}$ & $C_{Q_{1}}^{e}$ & $C_{Q_{2}}^{\mu}$ & $C_{Q_{2}}^{e}$ \\																
$	0.04	\pm	0.16	$ & $	-1.40	\pm	1.40	$ & $	-0.08	\pm	0.18	$ & $	-4.30	\pm	0.70	$ \\
\hline\hline																
$C_{Q_{1}}^{\prime \mu}$ & $C_{Q_{1}}^{\prime e}$ & $C_{Q_{2}}^{\prime \mu}$ & $C_{Q_{2}}^{\prime e}$ \\																
$	0.16	\pm	0.17	$ & $	-1.30	\pm	1.50	$ & $	-0.14	\pm	0.18	$ & $	-4.20	\pm	0.70	$ \\
\hline																
\end{tabular}\qquad
\begin{tabular}{|c|c|c|c|}
\hline																
\multicolumn{4}{|c|}{All observables  with $\chi^2_{\rm SM}=	 	215.1			$} \\											
\multicolumn{4}{|c|}{($\chi^2_{\rm min}=	 	151.3	;\; {\rm Pull}_{\rm SM}=	4.8	\sigma$)} \\											
\hline \hline																
\multicolumn{2}{|c|}{$\delta C_7$} &  \multicolumn{2}{c|}{$\delta C_8$}\\																
\multicolumn{2}{|c|}{$	0.06	\pm	0.03	$} & \multicolumn{2}{c|}{$	-0.80	\pm	0.40	$}\\								
\hline																
\multicolumn{2}{|c|}{$\delta C_7^\prime$} &  \multicolumn{2}{c|}{$\delta C_8^\prime$}\\																
\multicolumn{2}{|c|}{$	-0.01	\pm	0.02	$} & \multicolumn{2}{c|}{$	0.00	\pm	0.80	$}\\								
\hline																
$\delta C_{9}^{\mu}$ & $\delta C_{9}^{e}$ & $\delta C_{10}^{\mu}$ & $\delta C_{10}^{e}$ \\																
$	-1.20	\pm	0.18	$ & $	-6.70	\pm	1.20	$ & $	0.10	\pm	0.23	$ & $	4.00	\pm	5.00	$ \\
\hline\hline																
$\delta C_{9}^{\prime \mu}$ & $\delta C_{9}^{\prime e}$ & $\delta C_{10}^{\prime \mu}$ & $\delta C_{10}^{\prime e}$ \\																
$	0.07	\pm	0.33	$ & $	1.80	\pm	1.50	$ & $	-0.12	\pm	0.20	$ & $	0.00	\pm	5.00	$ \\
\hline\hline																
$C_{Q_{1}}^{\mu}$ & $C_{Q_{1}}^{e}$ & $C_{Q_{2}}^{\mu}$ & $C_{Q_{2}}^{e}$ \\																
$	0.07	\pm	0.10	$ & $	-1.40	\pm	0.90	$ & $	-0.06	\pm	0.13	$ & $	-4.20	\pm	1.50	$ \\
\hline\hline																
$C_{Q_{1}}^{\prime \mu}$ & $C_{Q_{1}}^{\prime e}$ & $C_{Q_{2}}^{\prime \mu}$ & $C_{Q_{2}}^{\prime e}$ \\																
$	0.19	\pm	0.10	$ & $	-1.30	\pm	1.90	$ & $	-0.13	\pm	0.14	$ & $	-4.20	\pm	0.90	$ \\
\hline																
\end{tabular}
}
\caption{Best fit values for the 20 operator fits to all observables, assuming 10\% error for the power corrections.
On the left (right) side we have the results excluding (including $B^+ \to K^{*+} \mu^+ \mu^-$ observables).
\label{tab:ALLwBplus_20D_C78910C12primes}} 
\end{center} 
\end{table}
%
%

In summary, the new data add another consistent piece to the New Physics interpretation of the various tensions in the $b \to s$ data.
Also the recent developments on the theoretical analyses of power corrections 
supports the New Physics interpretation.\\

{ {\bf Acknowledgements:} The work of T.H. was supported by  the  Cluster of  Excellence 
``Precision  Physics, Fundamental Interactions, and Structure of Matter'' 
(PRISMA$^+$ EXC 2118/1) funded by the German Research Foundation (DFG) 
within the German Excellence Strategy (Project ID 39083149), 
as well as BMBF Verbundprojekt 05H2018 - Belle II.  
T.H. thanks the CERN theory group for its hospitality during his regular visits to CERN. }


\providecommand{\href}[2]{#2}\begingroup\raggedright
 \endgroup

\end{document}